\documentclass[12pt,a4paper,british]{article}
\usepackage[T1]{fontenc}
\usepackage{babel}
\usepackage{graphics}
\usepackage{setspace}
\makeatletter

\providecommand{\LyX}{L\kern-.1667em\lower.25em\hbox{Y}\kern-.125emX\@}

 \newcommand{\lyxaddress}[1]{
   \par {\raggedright #1
   \vspace{1.4em}
   \noindent\par}
 }

\usepackage{amsfonts}
\usepackage{amsmath}

\DeclareTextCompositeCommand{\v}{OT1}{t}{%
    t\kern-.23em\raise.24ex\hbox{'}}

\newcommand{\C}{\ensuremath{\mathbb C}}
\newcommand{\R}{\ensuremath{\mathbb R}}
\newcommand{\Z}{\ensuremath{\mathbb Z}}
\newcommand{\G}{\ensuremath{\mathbb G}}

\newcommand{\N}{\ensuremath{\mathbb N}}

\newcommand{\NN}{\ensuremath{{\cal N}}}
\newcommand{\HH}{\ensuremath{{\cal H}}}

\newcommand{\bJ}{\ensuremath{\boldsymbol{J}}}
\newcommand{\bA}{\ensuremath{\boldsymbol{A}}}
\newcommand{\bzeta}{\ensuremath{\boldsymbol{\zeta}}}

\newcommand{\vr}{\varrho}
\newcommand{\dd}{\mathrm{d}}
\newcommand{\Ran}{\mathop{\mathrm{Ran}}}
\newcommand{\Coi}{C_0^\infty}
\newcommand{\ogsum}[1]{\sideset{}{^\oplus}\sum_{#1}}
\newcommand{\Fab}{ {}_2{F_1}}
\newcommand{\phna}{\phantom{\times\,\Biggl(}}

\newcommand{\Dom}{\mathop{\mathrm{Dom}}}
\DeclareMathOperator{\Imp}{Im}
\DeclareMathOperator{\Mat}{Mat}
\DeclareMathOperator{\rank}{rank}
\DeclareMathOperator{\Ker}{Ker}
\DeclareMathOperator{\diag}{diag}

\makeatother
\begin{document}

\title{Generalised boundary conditions for the~Aharonov-Bohm effect combined
with a~homogeneous magnetic field}

\date{}

\author{P. Exner\( ^{1,3} \), P. \v{S}\v{t}ov\'\i\v{c}ek\( ^{2,3} \),
P. Vyt\v{r}as\( ^{2} \)}

\maketitle

\lyxaddress{
  \( ^{1} \)Nuclear Physics Institute, Academy of Sciences,
  250 68 \v{R}e\v{z} near Prague, Czech Republic\\
  \( ^{2} \)Department of Mathematics, Faculty of Nuclear Science,
  Czech Technical University, Trojanova 13, 120 00 Prague, Czech
  Republic\\
  \( ^{3} \)Doppler Institute, Czech Technical University,
  B\v{r}ehov{\'a} 7, 115 19 Prague, Czech Republic
}

\begin{abstract}
  The most general admissible boundary conditions are derived for an
  idealised Aharonov-Bohm flux intersecting the plane at the origin on
  the background of a homogeneous magnetic field. A standard technique
  based on self-adjoint extensions yields a four-parameter family of
  boundary conditions; other two parameters of the model are the
  Aharonov-Bohm flux and the homogeneous magnetic field. The
  generalised boundary conditions may be regarded as a combination of
  the Aharonov-Bohm effect with a point interaction. Spectral
  properties of the derived Hamiltonians are studied in detail.
\end{abstract}

\section{Introduction}

The purpose of this paper is to determine the most general
admissible boundary conditions for the Aharonov-Bohm (AB) effect
in the plane on the background of a homogeneous magnetic field,
and also to investigate the basic properties of Hamiltonians
obtained this way. The history of the effect goes four decades
back and starts from the observation of Aharonov and Bohm
\cite{AB} that the behavior of a charged quantum particle is
influenced by a magnetic flux even if the field is zero in the
region where the particle is localized. A particularly elegant
treatment is possible in case of an idealized setup in which the
AB flux is concentrated along a line perpendicularly intersecting
the plane, conventionally at the origin \cite{Ruijsenaars}.

The boundary conditions of the last mentioned paper are not the
most general ones; the full family of such conditions giving the
AB effect in the plane was derived in \cite{DS}, and
simultaneously also in \cite{AT}. These generalised boundary
conditions may be interpreted as a combination of the AB effect
with a point interaction supported, too, at the origin, although
this is just one possible point of view. In any case they can be
described and investigated by the technique of self-adjoint
extensions which is in principle the same one as that used in the
paper \cite{Albeverio at al} in which two-dimensional point
interactions were introduced.

A natural question is what happens if such a system is placed into
a background homogeneous magnetic field. This problem attracted
some attention recently, even with a controversy: the papers
\cite{Thienel,Cavalcanti,HirokawaOgurisu} consider the ``pure'' AB
effect in this setting for the Pauli operator, i.e. a spin $1/2$
particle. The last named property leads to specific behavior
related to the Aharonov-Casher effect, which we will not discuss
here.

Our aim here is different: we are going to consider a spinless
particle with a point flux and a homogeneous background, and ask
about the most general class of boundary conditions analogous to
those of  \cite{DS,AT}. The basic difference between the
situations without and with a homogeneous magnetic field is that
in the former case the spectrum is absolutely continuous and equal
to the positive half-line possibly augmented with at most two
negative eigenvalues (depending on the choice of boundary
conditions) while in the latter case the spectrum is pure point
and the point flux and interaction gives rise to eigenvalues in
each gap between neighboring Landau levels. Our goal is to discuss
these spectral properties in detail.

\section{\label{Problem-prelim}Formulation of the problem, preliminaries}

We consider the symmetric operator
\[
L=-(\nabla -A(\nabla ))^{2},\textrm{ }\Dom (L)=
\Coi (\R ^{2}\setminus \{0\}),
\]
where the vector potential \( A \) is a sum of two parts,
\( A=A_{\mathrm{hmf}}+A_{\mathrm{AB}} \),
with the part \( A_{\mathrm{hmf}} \) corresponding to the homogeneous
magnetic field in the circular gauge,
\[
A_{\mathrm{hmf}}=-\frac{\imath B}{2}(-x_{2}dx_{1}+x_{1}dx_{2}),
\]
and with the part \( A_{\mathrm{AB}} \) corresponding to the idealised
AB effect,
\[
A_{\mathrm{AB}}=\frac{\imath \Phi }{2\pi r^{2}}
(-x_{2}dx_{1}+x_{1}dx_{2}),\textrm{ }r^{2}=x_{1}^{\, 2}+x_{2}^{\, 2}.
\]
Without loss of generality we may assume that \( B>0 \). Further,
we rescale the Aharonov-Bohm flux,
\[
\alpha =-\frac{\Phi }{2\pi },
\]
to have a variable which expresses the number of flux quanta
and, as usual, we make use of the gauge symmetry allowing us to assume
that \( \alpha \in \, ]0,1[\,  \). Hence the case $\Phi\in2\pi\Z$ is
excluded since it is gauge equivalent to the vanishing AB flux.
Our goal is to describe all the self-adjoint extensions of \( L \)
as well as to investigate their basic properties.

It is straightforward to determine the adjoint operator
\( L^{\ast } \),
\begin{eqnarray*}
  \psi \in \Dom (L^{\ast })\textrm{ } & \Longleftrightarrow  &
  \psi \in L^{2}(\R ^{2},\dd ^{2}x)\cap H^{2,2}_{\mathrm{loc}}
  (\R ^{2}\setminus \{0\})\\
  &  & \textrm{and }(\nabla -A(\nabla ))^{2}
  \psi \in L^{2}(\R ^{2},\dd ^{2}x).
\end{eqnarray*}
Next we can employ the rotational symmetry when using the polar coordinates
\( (r,\theta ) \) and decomposing the Hilbert space into the orthogonal
sum of the eigenspaces of the angular momentum,
\begin{equation}
  \label{ogsum}
  L^{2}(\R ^{2},\dd ^{2}x)=\ogsum {m\in \Z }
  L^{2}(\R _{+},r\, \dd r)\otimes \C \, e^{\imath m\theta }.
\end{equation}
In the polar coordinates the operator \( L \) (and correspondingly
\( L^{\ast } \)) takes the form
\[
L=-\frac{1}{r}\, \partial _{r}r\partial _{r}+
\frac{1}{r^{2}}\left( -\imath \partial _{\theta }+
  \alpha +\frac{Br^{2}}{2}\right) ^{2}.
\]
The operator \( L^{\ast } \) commutes on \( \Dom (L^{\ast }) \)
with the projectors \( P_{m} \) onto the eigenspaces of the angular
momentum,
\[
P_{m}\psi (r,\theta )=\frac{1}{2\pi }
\int _{0}^{2\pi }\psi (r,\theta')\,
e^{\imath m(\theta -\theta ')}\, \dd \theta ',
\]
and therefore \( L^{\ast } \) decomposes in correspondence with the
orthogonal sum (\ref{ogsum}),
\begin{equation}
  \label{decompL}
  L^{\ast }=\ogsum {m\in \Z }(L^{\ast })_{m}.
\end{equation}

Thus we can reduce the problem and work in the sectors \( \Ran P_{m} \),
\( m\in \Z  \). For a given spectral parameter \( \lambda \in \C  \)
we choose two independent solutions (except of particular values of
\( \lambda  \)) of the differential equation
\begin{equation}
  \label{eqm}
  \left( -\frac{1}{r}\, \partial _{r}r\partial _{r}+
    \frac{1}{r^{2}}\left( m+\alpha +
      \frac{Br^{2}}{2}\right) ^{2}\right) g(r)=\lambda \, g(r),
\end{equation}
namely
\begin{equation}\label{gm}
  \begin{aligned}
    g^{1}_{m}(\lambda;r) &= r^{|m+\alpha|}\,
    F\left(\beta(m,\lambda), \gamma(m), \frac{Br^2}{2}\right)
    \exp\left(-\frac{Br^2}{4}\right), \\
    g^{2}_{m}(\lambda;r) &= r^{|m+\alpha|}\,
    G\left(\beta(m,\lambda), \gamma(m), \frac{Br^2}{2}\right)
    \exp\left(-\frac{Br^2}{4}\right),
  \end{aligned}\end{equation}
where
\begin{equation}\label{beta}
  \begin{aligned}
    \beta(m,\lambda) &= \frac{1}{2}\left(1+m+\alpha+|m+\alpha|-
      \frac{\lambda}{B}\right),\\
    \gamma(m) &= 1+|m+\alpha|\,.
  \end{aligned}
\end{equation}
Here \( F \) and \( G \) are confluent hypergeometric
functions \cite[Chp. 13]{AbramowitzStegun},
\[
F(\beta ,\gamma ,z)=\sum ^{\infty }_{n=0}
\frac{(\beta )_{n}\, z^{n}}{(\gamma )_{n}\, n!}\, ,
\]
and
\begin{equation}
  \label{GF}
  G(\beta ,\gamma ,z)=\frac{\Gamma (1-\gamma )}
  {\Gamma (\beta -\gamma +1)}\, F(\beta ,\gamma ,z)+
  \frac{\Gamma (\gamma -1)}{\Gamma (\beta )}\,
  z^{1-\gamma }F(\beta -\gamma +1,2-\gamma ,z).
\end{equation}

Notice that \( F(\beta ,\gamma ,z) \) and \( G(\beta ,\gamma ,z) \)
are linearly dependent if and only if \( \beta \in -\Z _{+} \). Moreover,
\( F(\beta ,\gamma ,z) \) is an entire function, particularly, it
is regular at the origin while \( G(\beta ,\gamma ,z) \) has a singularity
there provided \( \gamma >1 \) and \( \beta \notin -\Z _{+} \),
and in that case it holds true that
\[
\lim _{z\to 0_{+}}z^{\gamma -1}G(\beta ,\gamma ,z)=
\frac{\Gamma (\gamma -1)}{\Gamma (\beta )}\, .
\]
Thus in the case when \( 1<\gamma <2 \) we have the asymptotic behaviour,
as \( z\to 0_{+} \),
\begin{equation}
  \label{Gasym}
  G(\beta ,\gamma ,z)=\frac{\Gamma (\gamma -1)}{\Gamma (\beta )}\,
  z^{1-\gamma }+\frac{\Gamma (1-\gamma )}{\Gamma (\beta -\gamma +1)}
  +O(z^{2-\gamma }).
\end{equation}

We shall also need some information about the asymptotic behaviour
at infinity. When \( z\to +\infty  \) it holds true that
\begin{equation}
  \label{ASF}
  F(\beta ,\gamma ,z)=\frac{\Gamma (\gamma )}
  {\Gamma (\gamma -\beta)}\, (-z)^{-\beta }
  \left( 1+O\left( z^{-1}\right) \right)
  +\frac{\Gamma (\gamma )}{\Gamma (\beta )}\,
  e^{z}z^{\beta -\gamma }\left( 1+O\left( z^{-1}\right) \right)
\end{equation}
and
\[
G(\beta ,\gamma ,z)=z^{-\beta }
\left( 1+O\left( z^{-1}\right) \right) .
\]

\section{\label{StandardHamilton}The standard Aharonov-Bohm Hamiltonian}

With the above preliminaries it is straightforward
to solve the spectral problem for the standard AB Hamiltonian
as we mentioned in the introduction.
This means to solve the eigenvalue problem
\[
L^{\ast }\psi =\lambda \psi
\]
with the boundary condition
\begin{equation}
  \label{BC}
  \lim _{r\to 0_{+}}\psi (r,\theta )=0.
\end{equation}
By virtue of the decomposition (\ref{decompL}) the problem is reduced
to the countable set of equations
\[
(L^{\ast })_{m}f=\lambda f,\textrm{ }m\in \Z ,
\]
and hence to the differential equations (\ref{eqm}).

The solution \( g^{2}_{m}(\lambda ;r) \) of (\ref{eqm}) is ruled
out because it contradicts the condition (\ref{BC}) and the solution
\( g^{1}_{m}(\lambda ;r) \) belongs to \( L^{2}(\R _{+},r\, \dd r) \)
if and only if \( \beta (m,\lambda )=-n \), with \( n\in \Z _{+} \).
Since it holds
\[
F(-n,1+\sigma ,z)=\frac{n!\, \Gamma (\sigma +1)}
{\Gamma (n+\sigma +1)}\, L_{n}^{\sigma }(z),\textrm{ }n\in \Z _{+},
\]
we get a countable set of eigenvalues,
\[
\lambda _{m,n}=B\, (m+\alpha +|m+\alpha |+2n+1),
\textrm{ }m\in \Z ,\textrm{ }n\in \Z _{+},
\]
with the corresponding eigenfunctions
\[
f_{m,n}(r,\theta )=C_{m,n}\, r^{|m+\alpha |}\, L_{n}^{|m+\alpha |}
\left( \frac{Br^{2}}{2}\right) \, \exp
\left( -\frac{Br^{2}}{4}\right) \, e^{\imath m\theta }
\]
where
\[
C_{m,n}=\left( \frac{B}{2}\right) ^{\frac{1}{2}(|m+\alpha |+1)}
\left( \frac{n!}{\pi \, \Gamma (n+|m+\alpha |+1)}\right) ^{1/2}
\]
are the normalisation constants.

As it is well known 
if we fix \( m\in \Z  \) then the functions
\( \{f_{m,n}(r,\theta )\}^{\infty }_{n=0} \)
form an orthonormal basis in
\( L^{2}(\R _{+},r\, \dd r)\otimes \C \, e^{\imath m\theta } \)
and so the complete set of eigenfunctions
\( \{f_{m,n}(r,\theta )\}_{m\in \Z \,,\textrm{ }n\in \Z _{+}} \)
is an orthonormal basis in
\( L^{2}(\R _{+},r\, \dd r)\otimes L^{2}([\, 0,2\pi \, ],\dd \theta ) \).
Since all the eigenvalues \( \lambda _{m,n} \) are real we get this
way a well defined self-adjoint operator which is an extension of
\( L \). We conventionally call it the standard AB Hamiltonian and
denote it by \( H^{AB} \). Thus the spectrum of \( H^{AB} \) is
pure point and can be written as a union of two parts,
\[
\sigma (H^{AB})=\sigma _{pp}(H^{AB})=
\{B(2k+1);\textrm{ }k\in \Z _{+}\}\cup
\{B(2\alpha +2k+1);\textrm{ }k\in \Z _{+}\}.
\]
Notice that the eigenvalues belonging to the first part are nothing
but the Landau levels. All the eigenvalues \( B(2k+1) \) have infinite
multiplicities while the multiplicity of the eigenvalue
\( B(2\alpha +2k+1) \)
is finite and equals \( k+1 \).

A final short remark concerning the Hamiltonian $H^{AB}$ is devoted to
the Green function. Naturally, the Green function is expressible as an
infinite series
\begin{displaymath}
  G^{AB}(z;r_1,\theta_1,r_2,\theta_2) = \frac{1}{2\pi}
  \sum_{m=-\infty}^\infty G_m^{AB}(z;r_1,r_2)\,
  e^{\imath m(\theta_1-\theta_2)}
\end{displaymath}
where
\begin{eqnarray*}
  G_m^{AB}(z;r_1,r_2) &=& 2\left(\frac{B}{2}\right)^{|m+\alpha|+1}
  (r_1r_2)^{|m+\alpha|}
  \exp\left(-\frac{1}{4}\,B(r_1^{\ 2}+r_2^{\ 2})\right)\\
  &&\times \sum_{n=0}^\infty\,
  \frac{n!}{\Gamma(n+|m+\alpha|+1)}\\
  &&\quad\times \frac{L_n^{|m+\alpha|}(\frac{1}{2}Br_1^{\ 2})\,
    L_n^{|m+\alpha|}(\frac{1}{2}Br_2^{\ 2})}
  {B(m+\alpha+|m+\alpha|+2n+1)-z}\,.
\end{eqnarray*}
The radial parts can be rewritten with the aid of the standard
construction of the Green function for ordinary differential operators
of second order,
\begin{eqnarray*}
  G_m^{AB}(z;r_1,r_2) &=& \left(\frac{B}{2}\right)^{|m+\alpha|+1}
  (r_1r_2)^{|m+\alpha|}
  \exp\left(-\frac{1}{4}\,B(r_1^{\ 2}+r_2^{\ 2})\right)\\
  &&\times\, \frac{\Gamma\big(-w(m,z)\big)}{\Gamma(|m+\alpha|+1)}\,
  F(-w(m,z),|m+\alpha|+1,r_<)\\
  &&\times\, G(-w(m,z),|m+\alpha|+1,r_>)
\end{eqnarray*}
where
\begin{displaymath}
  w(m,z) = \frac{z}{2B} - \frac{1}{2}(m+\alpha+|m+\alpha|+1)
\end{displaymath}
and $r_<=\min(r_1,r_2)$, $r_>=\max(r_1,r_2)$. This amounts to the
identity
\begin{eqnarray*}
  && \sum_{n=0}^\infty\,\frac{n!}{\Gamma(n+\sigma+1)}\,
  \frac{L_n^\sigma(y_1)\,L_n^\sigma(y_2)}{n-w} \\
  &&\qquad =\, \frac{\Gamma(-w)}{\Gamma(\sigma+1)}\,
  F(-w,\sigma+1,y_<)\,G(-w,\sigma+1,y_>).
\end{eqnarray*}
We do not expect that a simpler form for the Green function could be
derived since the Hamiltonian $H^{AB}$ enjoys only the rotational
symmetry.

\section{Self-adjoint extensions of \protect\( L\protect \)}

Recalling what has been summarised in Section \ref{Problem-prelim}
it is easy to determine the deficiency indices. The solution
\( g^{1}_{m}(\pm \imath ;r) \)
diverges exponentially at infinity (cf. (\ref{ASF})) while
\( g^{2}_{m}(\pm \imath ;r) \)
behaves well at infinity but has a singularity at the origin of the
order \( r^{-|m+\alpha |} \). Thus
\( g^{2}_{m}(\pm \imath ;r)\in L^{2}(\R _{+},r\, \dd r) \)
if and only if \( m=-1 \) or \( m=0 \). This means that the deficiency
indices are \( (2,2) \). For a basis in the deficiency subspaces
\( \NN _{\pm \imath } \) we can choose
\[
\{f_{m,\pm }(r,\theta )=
\frac{1}{\sqrt{2\pi }}\, N_{m}\, g^{2}_{m}(\pm \imath ;r)\,
e^{\imath m\theta };\textrm{ }m=-1,0\}.
\]
Thus
\begin{eqnarray*}
f_{-1,\pm }(r,\theta ) & = & \frac{1}{\sqrt{2\pi }}\,
N_{-1}\, r^{1-\alpha }\,
G\left( \frac{1}{2}\mp \frac{\imath }{2B},2-\alpha ,
\frac{Br^{2}}{2}\right) \, \exp
\left( -\frac{Br^{2}}{4}\right) \, e^{-\imath \theta },\\
f_{0,\pm }(r,\theta ) & = & \frac{1}{\sqrt{2\pi }}\, N_{0}\,
r^{\alpha }\, G\left( \frac{1}{2}+\alpha \mp \frac{\imath
    }{2B},1+\alpha ,
\frac{Br^{2}}{2}\right) \, \exp \left( -\frac{Br^{2}}{4}\right) ,
\end{eqnarray*}
where \( N_{-1} \) and \( N_{0} \) are normalisation constants making
the basis orthonormal.

We shall need the explicit values of \( N_{-1} \) and \( N_{0} \).
Using the relation
\[
W_{\upsilon ,\tau }(z)=z^{\tau +\frac{1}{2}}e^{-z/2}\,
G\left( \frac{1}{2}-\upsilon +\tau ,2\tau +1,z\right)
\]
where \( W \) is the Whittaker function we get
\begin{eqnarray*}
  N_{m}^{\ -2} & = &
  \int ^{\infty }_{0}|g^{2}_{m}(\pm \imath ;r)|^{2}\, r\dd r\\
  & = & \frac{1}{2}\left( \frac{2}{B}\right) ^{|m+\alpha |+1}
  \int _{0}^{\infty }x^{-1}W_{\varrho ,\sigma }(x)\,
  W_{\bar{\varrho },\sigma }(x)\, \dd x
\end{eqnarray*}
where
\[
\varrho =\frac{1}{2}\left( -m-\alpha +\frac{\imath }{B}\right) ,
\textrm{ }\sigma =\frac{1}{2}\, |m+\alpha |.
\]
Combining the identities \cite[2.19.24.6]{PBM}
\begin{eqnarray*}
  &  & \int ^{\infty }_{0}x^{-1}W_{\vr ,\sigma }(x)\,
  W_{\mu ,\sigma }(x)\, \dd x=\frac{\pi }{\sin (2\pi \sigma )}\\
  &  & \qquad \times \, \Biggl (-\frac{1}{\Gamma
    \left( \frac{1}{2}-\sigma -\mu \right) \, \Gamma
    \left( \frac{3}{2}+\sigma -\vr \right) }\,
  \Fab \biggl (\frac{1}{2}+\sigma -\mu ,1;\frac{3}{2}+
  \sigma -\vr ;1\biggr )\\
  &  & \qquad \phna +\frac{1}{\Gamma
    \left( \frac{1}{2}+\sigma -\mu \right)
    \, \Gamma \left( \frac{3}{2}-\sigma -\vr \right) }\,
  \Fab \biggl (\frac{1}{2}-\sigma -\mu ,1;
  \frac{3}{2}-\sigma -\vr ;1\biggr )\Biggr )
\end{eqnarray*}
and
\begin{eqnarray*}
  &  & \Fab (a,b;c;z)\, =\\
  &  & \qquad \frac{\Gamma (c)\, \Gamma (c-a-b)}
  {\Gamma (c-a)\, \Gamma (c-b)}\, \Fab (a,b;a+b-c+1;1-z)\\
  &  & \qquad +\, \frac{\Gamma (c)\,
   \Gamma (a+b-c)}{\Gamma (a)\, \Gamma (b)}\, (1-z)^{c-a-b}
 \Fab (c-a,c-b;c-a-b+1;1-z)
\end{eqnarray*}
we arrive at the relation\begin{eqnarray*}
  &  & \int ^{\infty }_{0}x^{-1}W_{\vr ,\sigma }(x)\,
  W_{\mu ,\sigma }(x)\, \dd x=
  \frac{\pi }{\sin (2\pi \sigma )(\mu -\vr )}\\
  &  & \qquad \times \left( -\frac{1}{\Gamma
      \left( \frac{1}{2}-\mu -\sigma \right) \, \Gamma
      \left( \frac{1}{2}-\vr +\sigma \right) }+
    \frac{1}{\Gamma \left( \frac{1}{2}-\mu +\sigma \right) \,
      \Gamma \left( \frac{1}{2}-\vr -\sigma \right) }\right) .
\end{eqnarray*}
Finally we get
\begin{eqnarray*}
  N_{-1} & = & \left( \frac{B}{2}\right) ^{\frac{1}{2}(1-\alpha )}
  \sqrt{\frac{\sin (\pi \alpha )}{2\pi }}\,
  \left( \Imp \frac{1}{\Gamma \left( -\frac{1}{2}+\alpha
        +\frac{\imath }{2B}\right) \, \Gamma
      \left( \frac{1}{2}-\frac{\imath }{2B}\right) }\right) ^{-1/2},\\
  N_{0} & = & \left( \frac{B}{2}\right) ^{\frac{1}{2}\alpha }
  \sqrt{\frac{\sin (\pi \alpha )}{2\pi }}\,
  \left( \Imp \frac{1}{\Gamma
      \left( \frac{1}{2}+\frac{\imath }{2B}\right) \,
      \Gamma \left( \frac{1}{2}+\alpha -
        \frac{\imath }{2B}\right) }\right) ^{-1/2}.
\end{eqnarray*}

Let us have a look at the asymptotic behaviour at the origin of the
basis functions in the deficiency subspaces \( \NN _{\pm \imath } \).
By (\ref{gm}) and (\ref{Gasym}) we have
\begin{equation}
  \label{AS}
  \begin{aligned}
    g^{2}_{-1}(\pm\imath;r) &= a_{-1,\pm}\,r^{-1+\alpha}+
    b_{-1,\pm}\,r^{1-\alpha}+O(r^{1+\alpha}), \\
    g^{2}_{0}(\pm\imath;r) &= a_{0,\pm}\,r^{-\alpha}+
    b_{0,\pm}\,r^{\alpha}+O(r^{2-\alpha}),
  \end{aligned}
\end{equation}
where
\begin{eqnarray*}
  a_{-1,\pm }=\frac{\Gamma (1-\alpha )}{\Gamma
    \left( \frac{1}{2}\mp \frac{\imath }{2B}\right) }\,
  \left( \frac{B}{2}\right) ^{-1+\alpha } & ,\textrm{ } &
  b_{-1,\pm }=\frac{\Gamma (-1+\alpha )}{\Gamma
    \left( -\frac{1}{2}+\alpha \mp \frac{\imath }{2B}\right) }\, ,\\
  a_{0,\pm }=\frac{\Gamma (\alpha )}{\Gamma \left( \frac{1}{2}+
      \alpha \mp \frac{\imath }{2B}\right) }\,
  \left( \frac{B}{2}\right) ^{-\alpha } & ,\textrm{ } &
  b_{0,\pm }=\frac{\Gamma (-\alpha )}
  {\Gamma \left( \frac{1}{2}\mp \frac{\imath }{2B}\right) }\, .
\end{eqnarray*}
The coefficients \( a_{m,\pm } \), \( b_{m,\pm } \) are related
to the normalisation constants \( N_{m} \) for it holds true that
\begin{equation}
  \label{DET}
  \det M_{-1}=-\frac{\imath }{1-\alpha }\,
  (N_{-1})^{-2},\textrm{ }\det M_{0}=-\frac{\imath }{\alpha }\,
  (N_{0})^{-2}.
\end{equation}
where
\[
M_{m}=\left(
  \begin{array}{cc}
    a_{m,+} & b_{m,+}\\
    a_{m,-} & b_{m,-}
  \end{array}\right) .
\]
Particularly, we shall need the fact that the matrices \( M_{-1} \)
and \( M_{0} \) are regular.

Let us now describe the closure of the operator \( L \). In virtue
of the decomposition (\ref{decompL}) we have\[
\bar{L}=\ogsum {m\in \Z }\bar{L}_{m}\]
where \( \bar{L}_{m}=(L^{\ast })_{m}^{\, \ast } \). As it is well known,
\( \psi \in \Dom (L^{\ast }) \) belongs to \( \Dom (\bar{L}) \)
if and only if
\( \left\langle \psi ,L^{\ast }\varphi \right\rangle =
\left\langle L^{\ast }\psi ,\varphi \right\rangle  \)
for all \( \varphi \in \NN _{\imath }+\NN _{-\imath } \). Thus
\( (L^{\ast })_{m}=\bar{L}_{m} \)
for \( m\neq \{-1,0\} \), and if \( m\in \{-1,0\} \) then
\( \varphi (r)\, e^{\imath m\theta }\in \Dom ((L^{\ast })_{m}) \)
belongs to \( \Dom (\bar{L}_{m}) \) if and only if
\[
\lim _{r\to 0_{+}}r\,
W(\overline{\varphi (r)},g^{2}_{m}(\pm \imath ,r))=0
\]
where \( W(f,g)=(\partial _{r}f)g-f\, \partial _{r}g \) is the Wronskian.
Using the asymptotic behaviour (\ref{AS}) and the regularity of the
matrix \( M_{m} \) we arrive at two conditions
\begin{eqnarray*}
  & \lim _{r\to 0_{+}}(-|m+\alpha |\, r^{-|m+\alpha |}
  \varphi (r)-r^{-|m+\alpha |+1}\partial _{r}\varphi (r))=0, & \\
  & \lim _{r\to 0_{+}}(|m+\alpha |\, r^{|m+\alpha |}
 \varphi (r)-r^{|m+\alpha |+1}\partial _{r}\varphi (r))=0, &
\end{eqnarray*}
which can be rewritten in the equivalent form,
\[
\lim _{r\to 0_{+}}r^{-2|m+\alpha |+1}\partial _{r}(r^{|m+\alpha |}
\varphi (r))=0,\textrm{ }\lim _{r\to 0_{+}}r^{|m+\alpha |}\varphi
(r)=0.
\]
But since
\[
r^{-|m+\alpha |}|\varphi (r)|\leq \frac{1}{2|m+\alpha |}\,
\sup _{x\in \, ]0,r[}|x^{-2|m+\alpha |+1}
\partial _{x}(x^{|m+\alpha |}\varphi (x))|
\]
we finally get a sufficient and necessary condition for
\( \varphi (r)\, e^{\imath m\theta }\in \Dom ((L^{\ast })_{m}) \)
to belong to \( \Dom (\bar{L}) \), namely
\begin{equation}
  \label{CL}
  \begin{aligned}
    {}&\lim_{r\to 0_+}r^{-1+\alpha}\varphi(r)=0
    \text{ and } \lim_{r\to 0_+}r^{\alpha}\varphi'(r)=0
    &\quad\text{if }& m=-1, \\
    {}&\lim_{r\to 0_+}r^{-\alpha}\varphi(r)=0 \text{ and }
    \lim_{r\to 0_+}r^{-\alpha+1}\varphi'(r)=0
    &\quad\text{if }& m=0.
  \end{aligned}
\end{equation}

This shows that if
\( \psi \in \Dom (L^{\ast })=
\Dom (\bar{L})+\NN _{\imath }+\NN _{-\imath } \)
then
\begin{eqnarray*}
  \psi (r,\theta ) & = & \left( \Phi ^{1}_{1}(\psi )
    r^{-1+\alpha }+\Phi ^{1}_{2}(\psi )r^{1-\alpha }\right)
  e^{-\imath \theta }+\Phi ^{2}_{1}(\psi )r^{-\alpha }+
  \Phi ^{2}_{2}(\psi )r^{\alpha }\\
  &  & +\textrm{ a regular part}.
\end{eqnarray*}
Let us formally introduce the functionals \( \Phi _{j}^{k} \) on
\( \Dom (L^{\ast }) \),
\begin{eqnarray*}
  \Phi ^{-1}_{1}(\psi ) & = & \lim _{r\to 0_{+}}r^{1-\alpha }
  \frac{1}{2\pi }\int ^{2\pi }_{0}\psi (r,\theta )\,
  e^{\imath \theta }\dd \theta ,\\
  \Phi ^{-1}_{2}(\psi ) & = & \lim _{r\to 0_{+}}r^{-1+\alpha }
  \left( \frac{1}{2\pi }\int ^{2\pi }_{0}\psi (r,\theta )\,
    e^{\imath \theta }\dd \theta -\Phi ^{1}_{1}(\psi )\,
    r^{-1+\alpha }\right) ,\\
  \Phi ^{0}_{1}(\psi ) & = & \lim _{r\to 0_{+}}r^{\alpha }
  \frac{1}{2\pi }\int ^{2\pi }_{0}\psi (r,\theta )\dd \theta ,\\
  \Phi ^{0}_{2}(\psi ) & = & \lim _{r\to 0_{+}}r^{-\alpha }
  \left( \frac{1}{2\pi }\int ^{2\pi }_{0}\psi (r,\theta )\dd \theta
    -\Phi ^{2}_{1}(\psi )\, r^{-\alpha }\right) .
\end{eqnarray*}
Notice that the upper index refers to the sector of angular momentum
while the lower index refers to the order of the singularity. If
\( \psi \in \Dom (\bar{L}) \)
then according to (\ref{CL}) it actually holds
\( \Phi ^{k}_{j}(\psi )=0 \)
for \( j=1,2 \), \( k=-1,0 \). On the other hand, if
\( \psi \in \NN _{\imath }+\NN _{-\imath } \)
and \( \Phi ^{k}_{j}(\psi )=0 \) for all indices
\( j=1,2 \), \( k=-1,0 \),
then \( \psi =0 \) (this is again guaranteed by the regularity of
the matrices \( M_{-1} \) and \( M_{0} \)).

Let us introduce some more notation. It is convenient to arrange the
functionals \( \Phi ^{k}_{j} \) into column vectors as follows,\[
\Phi _{j}(\psi )=\left( \begin{array}{c}
\Phi ^{-1}_{j}(\psi )\\
\Phi ^{0}_{j}(\psi )
\end{array}\right) ,\textrm{ }j=1,2.\]
Further, applying the functionals to the basis functions in
\( \NN _{\imath }+\NN _{-\imath } \)
we obtain four \( 2\times 2 \) diagonal matrices. More precisely,
set
\[
\left( \Phi _{j,\pm }\right) _{k\ell }=
\sqrt{2\pi }\, \Phi _{j}^{k-2}(f_{\ell -2,\pm }),\textrm{ }j,k,\ell
=1,2.
\]
Then
\[
\Phi _{1,\pm }=\left(
  \begin{array}{cc}
    N_{-1}a_{-1,\pm } & 0\\
    0 & N_{0}\, a_{0,\pm }
  \end{array}\right) ,
\textrm{ }\Phi _{2,\pm }=\left(
  \begin{array}{cc}
    N_{-1}b_{-1,\pm } & 0\\
    0 & N_{0}\, b_{0,\pm }
  \end{array}\right) .
\]

Now it is straightforward to give a formal definition of a self-adjoint
extension \( H^{U} \) of the symmetric operator \( L \) determined
by a unitary operator \( U:\NN _{\imath }\to \NN _{-\imath } \).
We identify \( U \) with a unitary \( 2\times 2 \) matrix via the
choice of the orthonormal bases \( \{f_{-1,\pm },f_{0,\pm }\} \)
in \( \NN _{\pm \imath } \).The self-adjoint operator \( H^{U} \)
is unambiguously defined by the condition: \( H^{U}\subset L^{\ast } \)
and \( \psi \in \Dom (L^{\ast }) \) belongs to \( \Dom (H^{U}) \)
if and only if\begin{equation}
\label{RAN}
\left( \begin{array}{c}
\Phi _{1}(\psi )\\
\Phi _{2}(\psi )
\end{array}\right) \in \Ran \left( \begin{array}{c}
\Phi _{1,+}+\Phi _{1,-}U\\
\Phi _{2,+}+\Phi _{2,-}U
\end{array}\right) .
\end{equation}
However condition (\ref{RAN}) is rather inconvenient and we shall
replace it in the next section by another one which is more suitable
for practical purposes.

\section{Boundary conditions}

To turn (\ref{RAN}) into a convenient requirement which would involve
boundary conditions we shall need the following proposition. Set
\[
D=\left( \begin{array}{cc}
    1-\alpha  & 0\\
    0 & \alpha
  \end{array}\right) .
\]
\emph{There is a one-to-one correspondence between unitary matrices
\( U\in U(2) \) and couples of matrices \( X_{1},X_{2}\in \Mat (2,\C ) \)
obeying\begin{equation}
\label{two}
\rank \left( \begin{array}{c}
X_{1}\\
X_{2}
\end{array}\right) =2
\end{equation}
and\begin{equation}
\label{RE}
X_{1}^{\, \ast }DX_{2}=X_{2}^{\, \ast }DX_{1}
\end{equation}
modulo the right action of the group of regular matrices \( GL(2,\C ) \).
The one-to-one correspondence is given by the equality}\[
\Ran \left( \begin{array}{c}
X_{1}\\
X_{2}
\end{array}\right) \in \Ran \left( \begin{array}{c}
\Phi _{1,+}+\Phi _{1,-}U\\
\Phi _{2,+}+\Phi _{2,-}U
\end{array}\right) .\]

Let us note that the equivalence class of a couple \( (X_{1},X_{2}) \)
modulo \( GL(2,\C ) \) corresponds to a two-dimensional subspace
in \( \C ^{4} \) and hence to a point in the Grassmann manifold
\( \G _{2}(\C ^{4}) \).
The complex dimension of \( \G _{2}(\C ^{4}) \) equals 4, i.e.
\( \dim _{\R }\G _{2}(\C ^{4})=8 \).
The points of \( \G _{2}(\C ^{4}) \) obeying the ({}``real'') condition
(\ref{RE}) form a real 4-dimensional submanifold which is diffeomorphic,
according to the proposition, to the unitary group \( U(2) \).

To verify the proposition we first show that to any couple
\( (X_{1},X_{2}) \)
with the properties (\ref{two}), (\ref{RE}) there are related unique
\( Y\in GL(2,\C ) \) and \( U\in U(2) \) such that
\begin{equation}
  \label{YU}
  \left( \begin{array}{c}
      X_{1}\\
      X_{2}
    \end{array}\right) Y=\bJ \left( \begin{array}{c}
      I\\
      U
    \end{array}\right)
\end{equation}
where we have set\[
\bJ =\left( \begin{array}{cc}
\Phi _{1,+} & \Phi _{1,-}\\
\Phi _{2,+} & \Phi _{2,-}
\end{array}\right) =\left( \begin{array}{cccc}
N_{-1}a_{-1,+} & 0 & N_{-1}a_{-1,-} & 0\\
0 & N_{0}a_{0,+} & 0 & N_{0}a_{0,-}\\
N_{-1}b_{-1,+} & 0 & N_{-1}b_{-1,-} & 0\\
0 & N_{0}b_{0,+} & 0 & N_{0}b_{0,-}
\end{array}\right) .
\]
Using (\ref{DET}) one easily finds that \( \bJ  \) is regular and
\[
\bJ ^{-1}=\imath \left( \begin{array}{cc}
    D & 0\\
    0 & D
  \end{array}\right) \left(
  \begin{array}{cc}
    \Phi _{2,-} & -\Phi _{1,-}\\
    -\Phi _{2,+} & \Phi _{1,+}
  \end{array}\right) .
\]
Let us introduce another couple of matrices,
\( V_{+},V_{-}\in \Mat (2,\C ) \),
by the relation\[
\left( \begin{array}{c}
V_{-}\\
V_{+}
\end{array}\right) =\bJ ^{-1}\left( \begin{array}{c}
X_{1}\\
X_{2}
\end{array}\right) ,
\]
thus
\( V_{\pm }=\mp \, \imath D(\Phi _{2,\pm }X_{1}-\Phi _{1,\pm }X_{2}) \).
It follows that
\[
V^{\, \ast }_{\pm }V_{\pm }=\left(
  \begin{array}{cc}
    X^{\, \ast }_{1} & X^{\, \ast }_{2}
  \end{array}\right) \left(
  \begin{array}{cc}
    \Phi _{2,\pm }^{\, \ast }D^{2}\Phi _{2,\pm } &
    -\Phi ^{\, \ast }_{2,\pm }D^{2}\Phi _{1,\pm }\\
    -\Phi ^{\, \ast }_{1,\pm }D^{2}\Phi _{2,\pm } &
    \Phi ^{\, \ast }_{1,\pm }D^{2}\Phi _{1,\pm }
  \end{array}\right) \left( \begin{array}{c}
    X_{1}\\
    X_{2}
  \end{array}\right)
\]
and, consequently,
\[
V^{\, \ast }_{-}V_{-}-V^{\, \ast }_{+}V_{+}=\left(
  \begin{array}{cc}
    X^{\, \ast }_{1} & X^{\, \ast }_{2}
  \end{array}\right) \left(
  \begin{array}{cc}
    0 & -\imath D\\
    \imath D & 0
  \end{array}\right) \left(
  \begin{array}{c}
    X_{1}\\
    X_{2}
  \end{array}\right) =
\imath (X^{\, \ast }_{2}DX_{1}-X^{\, \ast }_{1}DX_{2})
\]
for \( \Phi _{j,\pm } \) and \( D \) commute (all of them are diagonal),
\( \Phi ^{\, \ast }_{j,\pm }=\Phi _{j,\mp } \) and\[
-\Phi _{1,+}\Phi _{2,-}+\Phi _{1,-}\Phi _{2,+}=\imath D^{-1}\]
(cf. (\ref{DET})). Owing to the property (\ref{RE}) we have
\begin{equation}
  \label{VV}
  V^{\, \ast }_{-}V_{-}=V^{\, \ast }_{+}V_{+}
\end{equation}
which jointly with the property (\ref{two}) implies that
\[
\Ker V_{-}=\Ker V_{+}=\Ker \left(
  \begin{array}{c}
    V_{-}\\
    V_{+}
  \end{array}\right) =\Ker \left(
  \begin{array}{c}
    X_{1}\\
    X_{2}
  \end{array}\right) =0.
\]
The only possible choice of the matrices \( Y \) and \( U \) satisfying
(\ref{YU}) is\[
Y=V^{\, -1}_{-},\textrm{ }U=V_{+}V^{\, -1}_{-}.\]
The matrix \( U \) is actually unitary because of (\ref{VV}).

Conversely, we have to show that any couple of matrices \( X_{1} \),
\( X_{2} \) related to a unitary matrix \( U \) according to the
rule\[
\left( \begin{array}{c}
X_{1}\\
X_{2}
\end{array}\right) =\bJ \left( \begin{array}{c}
I\\
U
\end{array}\right) \]
obeys (\ref{two}) and (\ref{RE}). Condition (\ref{two}) is obvious
since \( \bJ  \) is regular and condition (\ref{RE}) is again a
matter of a direct computation. In more detail, since it holds\[
X^{\, \ast }_{1}DX_{2}-X_{2}^{\, \ast }DX_{1}=\left( \begin{array}{cc}
I & U^{\ast }
\end{array}\right) \bJ ^{\ast }\left( \begin{array}{cc}
0 & D\\
-D & 0
\end{array}\right) \bJ \left( \begin{array}{c}
I\\
U
\end{array}\right) \]
it suffices to verify that
\[
\bJ ^{\ast }\left( \begin{array}{cc}
    0 & D\\
    -D & 0
  \end{array}\right) \bJ =\imath \left( \begin{array}{cc}
    I & 0\\
    0 & -I
  \end{array}\right) .
\]
This concludes the proof of the above proposition.

Using this correspondence one can relate to a couple
\( X_{1},X_{2}\in \Mat (2,\C ) \)
obeying (\ref{two}) and (\ref{RE}) a self-adjoint extension \( H \)
determined by the condition
\begin{equation}
  \label{DOM}
  \psi \in \Dom (H)\Longleftrightarrow \left( \begin{array}{c}
      \Phi _{1}(\psi )\\
      \Phi _{2}(\psi )
    \end{array}\right) \in \Ran \left( \begin{array}{c}
      X_{1}\\
      X_{2}
    \end{array}\right) .
\end{equation}
Two couples \( (X_{1},X_{2}) \) and \( (X'_{1},X'_{2}) \) determine
the same self-adjoint extension if and only if there exists a regular
matrix \( Y \) such that \( (X'_{1},X'_{2})=(X_{1}Y,X_{2}Y) \).
Moreover, all the self-adjoint extensions can be obtained in this
way.

We shall restrict ourselves to an open dense subset in the space of
all self-adjoint extensions by requiring the matrix \( X_{2} \) to
be regular. In that case we can set directly \( X_{2}=I \) and rename
\( X_{1}=\Lambda  \). Thus \( \Lambda  \) is a \( 2\times 2 \)
complex matrix satisfying
\begin{equation}
  \label{DLambda}
  D\Lambda =\Lambda ^{\ast }D.
\end{equation}
The corresponding self-adjoint extension will be denoted
\( H^{\Lambda } \).
The condition (\ref{DOM}) simplifies in an obvious way. We conclude
that \( H^{\Lambda }\subset L^{\ast } \) and
\( \psi \in \Dom (L^{\ast }) \)
belongs to \( \Dom (H^{\Lambda }) \) if and only if
\begin{equation}
  \label{BCfinal}
  \Phi _{1}(\psi )=\Lambda \Phi _{2}(\psi ),
\end{equation}
and this is in fact the sought boundary condition.

Matrices \( \Lambda  \) obeying (\ref{DLambda}) can be parametrised
by four real parameters (or two real and one complex). We choose the
parameterisation\[
\Lambda =\left( \begin{array}{cc}
u & \alpha \bar{w}\\
(1-\alpha )w & v
\end{array}\right) ,\textrm{ }u,v\in \R ,\textrm{ }w\in \C .\]
The relation between \( \Lambda  \) and \( U \) reads\begin{equation}
\label{LambdaU}
\Lambda =(\Phi _{1,+}+\Phi _{1,-}U)(\Phi _{2,+}+\Phi _{2,-}U)^{-1}
\end{equation}
(provided the RHS makes sense).

The {}``most regular'' among the boundary conditions is
\( \Phi _{1}(\psi )=0 \),
i.e. the one determined by \( \Lambda =0 \), and the corresponding
self-adjoint extension is nothing but the standard Aharonov-Bohm
Hamiltonian \( H^{AB} \) discussed in Section \ref{StandardHamilton}.
According to (\ref{LambdaU}) \( H^{AB} \) corresponds to
the unitary matrix
\[
U=-\Phi ^{\, -1}_{1,-}\Phi _{1,+}=
\diag \left\{ -\frac{\Gamma \left( \frac{1}{2}+
\frac{\imath }{2B}\right) }
{\Gamma \left( \frac{1}{2}-\frac{\imath }{2B}\right) },-
\frac{\Gamma \left( \frac{1}{2}+\alpha +
\frac{\imath }{2B}\right) }{\Gamma
\left( \frac{1}{2}+\alpha -\frac{\imath }{2B}\right) }\right\} .
\]

\section{The spectrum}

Let us now proceed to the discussion of spectral properties of the
described self-adjoint extensions. It is clear from what has been
explained up to now that everything interesting is happening in the
two critical sectors of the angular momentum labeled by \( m=-1 \)
and \( m=0 \). To state it more formally we decompose the Hilbert
space into an orthogonal sum of the {}``stable'' and {}``critical''
parts,
\[
\HH =\HH _{s}\oplus \HH _{c}
\]
where
\[
\HH _{s}=\ogsum {m\in \Z \setminus
  \{-1,0\}}L^{2}(\R _{+},r\, \dd r)\otimes \C \,
e^{\imath m\theta },\textrm{ }\HH _{c}=L^{2}(\R _{+},r\, \dd r)
\otimes (\C \, e^{-\imath \theta }\oplus \C \, 1).
\]
A self-adjoint extension \( H^{\Lambda } \) decomposes
correspondingly,
\[
H^{\Lambda }=H^{\Lambda }|_{\HH _{s}}
\oplus H^{\Lambda }|_{\HH _{c}}\,,
\]
and we know that on \( \HH _{s} \) the operator \( H^{\Lambda } \)
coincides with the standard AB Hamiltonian,\[
H^{\Lambda }|_{\HH _{s}}=H^{AB}|_{\HH _{s}}.\]

Thus
\[
\sigma (H^{\Lambda })=\sigma (H^{AB}|_{\HH _{s}})\cup
\sigma (H^{\Lambda }|_{\HH _{c}})
\]
and, as explained in Section \ref{StandardHamilton},
\[
\sigma (H^{AB}|_{\HH _{s}})=\{B(2k+1);
\textrm{ }k\in \Z _{+}\}\cup \{B(2k+2\alpha +1);\textrm{ }k\in \N \}
\]
where the multiplicity of the eigenvalue \( B(2k+1) \) is infinite
while the multiplicity of the eigenvalue \( B(2k+2\alpha +1) \) equals
\( k \). On the other hand,
\[
\sigma (H^{AB}|_{\HH _{c}})=\{B(2k+1);
\textrm{ }k\in \Z _{+}\}\cup \{B(2k+2\alpha +1);
\textrm{ }k\in \Z_{+}\}
\]
where all the eigenvalues are simple (the first set is a contribution
of the sector \( m=-1 \) while the second one comes from the sector
\( m=0 \)). Since the deficiency indices are finite the Krein's formula
jointly with Weyl Theorem \cite[Theorem XIII.14]{ReedSimon4} tells
us that the essential spectrum
\( \sigma _{ess}(H^{\Lambda }|_{\HH _{c}}) \)
is empty for any \( \Lambda  \). Thus the spectrum of
\( H^{\Lambda }|_{\HH _{c}} \)
is formed by eigenvalues which are at most finitely degenerated and
have no finite accumulation points.

Let us derive the equation on eigenvalues for the restriction
\( H^{\Lambda }|_{\HH _{c}} \).
Let \( \lambda \in \R  \). In each of the sectors \( m=-1,0 \) there
exists exactly one (up to a multiplicative constant) solution of the
equation \( (L^{\ast })_{m}f=\lambda f \) which is \( L^{2} \)-integrable
at infinity (with respect to the measure \( r\, \dd r \)) and we
may take for it the function
\( g^{2}_{m}(\lambda ;r)\, e^{\imath m\theta } \)
(cf. (\ref{gm})). For a second linearly independent solution one
may take \( g^{1}_{m}(\lambda ;r)\, e^{\imath m\theta } \) provided
\( \beta (m,\lambda )\not \in -\Z _{+} \) (cf. (\ref{beta})). If
\( \beta (m,\lambda )\in -\Z _{+} \) then a possible choice of a
second linearly independent solution is
\[
r^{|m+\alpha |}\, H\biggl (\beta (m,\lambda ),\gamma
(m),\frac{Br^{2}}{2}\biggr )\, \exp \left( -\frac{Br^{2}}{4}\right)
\]
where\[
H(\beta ,\gamma ,z)=z^{1-\gamma }F(\beta -\gamma +1,2-\gamma ,z)\]
(cf. (\ref{GF})).

Thus \( \lambda  \) is an eigenvalue of \( H^{\Lambda }|_{\HH _{c}} \)
if and only if there exists a vector
\( (\mu ,\nu )\in \C ^{2}\setminus \{0\} \)
such that the function
\[
\psi _{\lambda }(r,\theta )=\mu \, g^{2}_{-1}(\lambda ;r)\,
e^{-\imath \theta }+\nu \, g^{2}_{0}(\lambda ;r)
\]
satisfies the boundary condition (\ref{BCfinal}). Using again (\ref{gm})
and (\ref{Gasym}) one finds that
\[
\Phi _{1}(\psi _{\lambda })=\left( \begin{array}{cc}
    a_{-1} & 0\\
    0 & a_{0}
  \end{array}\right) \left( \begin{array}{c}
    \mu \\
    \nu
  \end{array}\right) ,\textrm{ }\Phi _{2}(\psi _{\lambda })=
\left( \begin{array}{cc}
    b_{-1} & 0\\
    0 & b_{0}
  \end{array}\right) \left( \begin{array}{c}
    \mu \\
    \nu
  \end{array}\right) ,
\]
where
\begin{eqnarray*}
  a_{-1}=\frac{\Gamma (1-\alpha )}
  {\Gamma \left( \frac{1}{2}-\frac{\lambda }{2B}\right) }\,
  \left( \frac{B}{2}\right) ^{-1+\alpha } & ,\textrm{ } & b_{-1}
  =\frac{\Gamma (-1+\alpha )}{\Gamma
    \left( -\frac{1}{2}+\alpha -\frac{\lambda }{2B}\right) }\, ,\\
  a_{0}=\frac{\Gamma (\alpha )}{\Gamma \left(
      \frac{1}{2}+\alpha -\frac{\lambda }{2B}\right) }\,
  \left( \frac{B}{2}\right) ^{-\alpha } & ,\textrm{ } & b_{0}=
  \frac{\Gamma (-\alpha )}{\Gamma
    \left( \frac{1}{2}-\frac{\lambda }{2B}\right) }\, .
\end{eqnarray*}
This immediately leads to the desired equation on eigenvalues which
takes the form \( \det \bA =0 \) where\[
\bA =\left( \begin{array}{cc}
a_{-1} & 0\\
0 & a_{0}
\end{array}\right) \left( \begin{array}{c}
\mu \\
\nu
\end{array}\right) -\Lambda \left( \begin{array}{cc}
b_{-1} & 0\\
0 & b_{0}
\end{array}\right) .\]
After the substitution
\[
z=\frac{1}{2}-\frac{\lambda }{2B},\textrm{ i}.e.
\textrm{ }\lambda =B(1-2z),
\]
we get
\begin{eqnarray*}
  &  & \frac{\Gamma (1-\alpha )\Gamma (\alpha )}
  {\Gamma (z)\Gamma (z+\alpha )}\, \frac{2}{B}-
  \frac{\Gamma (\alpha )\Gamma (\alpha -1)}
  {\Gamma (z+\alpha -1)\Gamma (z+\alpha )}
  \left( \frac{2}{B}\right) ^{\alpha }u\\
  &  & \qquad -\, \frac{\Gamma (1-\alpha )\Gamma (-\alpha )}
  {\Gamma (z)^{2}}\left( \frac{2}{B}\right) ^{1-\alpha }v\\
  &  & \qquad +\, \frac{\Gamma (\alpha -1)\Gamma (-\alpha )}
  {\Gamma (z)\Gamma (z+\alpha -1)}\, (uv-\alpha (1-\alpha )|w|^{2})\,
  =\, 0.
\end{eqnarray*}
To simplify somewhat the form of the equation it is convenient to
rescale the parameters as follows,
\begin{equation}
  \label{SUBS}
  \xi =\left( \frac{B}{2}\right) ^{1-\alpha }
  \frac{\Gamma (\alpha )}{\Gamma (2-\alpha )}\,
  u,\textrm{ }\eta =\left( \frac{B}{2}\right) ^{\alpha }
  \frac{\Gamma (1-\alpha )}{\Gamma (1+\alpha )}\,
  v,\textrm{ }\zeta =\sqrt{\frac{B}{2}}\, |w|.
\end{equation}
Finally we arrive at an equation depending on three real parameters
\( \xi ,\eta ,\zeta  \), namely
\begin{equation}
  \label{EV}
  \frac{1}{\Gamma (z)\, \Gamma (z+\alpha )}+
  \frac{\xi }{\Gamma (z+\alpha -1)\, \Gamma (z+\alpha )}+
  \frac{\eta }{\Gamma (z)^{2}}+
  \frac{\xi \, \eta -\zeta ^{2}}{\Gamma (z)\, \Gamma (z+\alpha -1)}=0.
\end{equation}

There is no chance to solve equation (\ref{EV}) explicitly apart
of some particular cases. One of them, of course, corresponds to the
standard AB Hamiltonian. This case is determined by the values of
parameters \( \xi =\eta =\zeta =0 \) and the roots of (\ref{EV})
form the set \( -\Z _{+}\cup (-\alpha -\Z _{+}) \). Consider also
the case when when \( \xi =\eta =0 \) and \( \zeta \neq 0 \) with
the set of roots equal to
\( -\Z _{+}\cup (-\alpha -\Z _{+})\cup \{1-\alpha +\zeta ^{-2}\} \).
Comparing the latter case to the former one we see that there is one
additional root, namely \( 1-\alpha +\zeta ^{-2} \), which escapes
to infinity when \( \zeta \to 0 \).

In the last particular case one can also consider the limit
\( \zeta \to \infty  \).
More generally, suppose that \( \det \Lambda \neq 0 \), i.e.
\( \xi \eta -\zeta ^{2}\neq 0 \),
replace \( \Lambda  \) with \( t\, \Lambda  \) in (\ref{BCfinal})
and take the limit \( t\to \infty  \). The limiting boundary condition
reads\[
\Phi _{2}(\psi )=0\]
and the corresponding self-adjoint extension which we shall call
\( H^{\infty } \)
is one of those omitted when we restricted ourselves to an open dense
subset in the space of all self-adjoint extensions (regarded as a
a 4-dimensional real manifold). Equation (\ref{EV}) reduces in this
limit to the equation\begin{equation}
\label{RINF}
\frac{1}{\Gamma (z)\, \Gamma (z+\alpha -1)}=0
\end{equation}
with the set of roots \( -\Z _{+}\cup (1-\alpha -\Z _{+}). \)

Another case when equation (\ref{EV}) simplifies though it is not
solvable explicitly is \( \zeta =0 \). This is easy to understand
since if \( \zeta =0 \) then the matrix \( \Lambda  \) is diagonal
and the two critical sectors of angular momentum do not interfere.
This is reflected in the fact that the equation (\ref{EV}) splits
into two independent equations,
\[
\frac{1}{\Gamma (z)}+\frac{\xi }{\Gamma (z+\alpha -1)}=0,
\textrm{ }\frac{1}{\Gamma (z+\alpha )}+\frac{\eta }{\Gamma (z)}=0.
\]

Let us shortly discuss the dependence of roots of equation (\ref{EV})
on the parameters \( \xi ,\eta ,\zeta  \). Since the derivative of
the LHS of (\ref{EV}) with respect to \( z \) and with the values
of parameters \( (\xi ,\eta ,\zeta )=(0,0,0) \) equals
\[
\frac{(-1)^{m}m!}{\Gamma (-m+\alpha )}\neq 0\textrm{ for
  }z=-m,\textrm{ and }\frac{(-1)^{m}m!}{\Gamma (-m-\alpha )}\neq
0\textrm{ for }z=-m-\alpha ,
\]
where \( m\in \Z _{+} \), the standard Implicit Function Theorem
(analytic case) is sufficient to conclude that the roots are analytic
functions in \( \xi ,\eta ,\zeta  \) at least in some neighbourhood
of the origin (depending in general on the root). Let us denote by
\( z_{1,m}(\xi ,\eta ,\zeta ) \) and \( z_{2,m}(\xi ,\eta ,\zeta ) \)
the roots of (\ref{EV}) regarded as analytic functions in
\( \xi ,\eta ,\zeta  \)
and such that \( z_{1,m}(0,0,0)=-m \) and \( z_{2,m}(0,0,0)=-\alpha -m \),
with \( m\in \Z _{+} \). A straightforward computation results in
the following power series truncated at degree 4.

Set
\begin{eqnarray*}
  h^{0}_{m}(z) & = & \sum ^{m}_{j=1}\frac{1}{j}-\gamma -\psi (z),\\
  h^{1}_{m}(z) & = & \frac{\pi ^{2}}{6}+
  \sum ^{m}_{j=1}\frac{1}{j^{2}}-\psi '(z),\\
  h^{2}_{m}(z) & = & -2\, \bzeta (3)+
  2\sum ^{m}_{j=1}\frac{1}{j^{3}}-\psi ''(z),
\end{eqnarray*}
where \( \gamma  \) is the Euler constant,
\( \psi (z)=\Gamma '(z)/\Gamma (z) \)
is the digamma function and \( \bzeta  \) is the zeta function. Then
\begin{eqnarray}
  z_{1,m}(\xi ,\eta ,\zeta ) & = &
  -m+\frac{\left( -1\right) ^{m+1}}{m!\,
    \Gamma (-1-m+\alpha )}\, \xi +
  \frac{h_{m}^{0}(-1-m+\alpha )}{(m!)^{2}\, \Gamma (-1-m+\alpha
    )^{2}}\,
  \xi ^{2}\nonumber \\
  &  & +\, \frac{\left( -1\right) ^{m+1}
    \left( 3\, h_{m}^{0}(-1-m+\alpha )^{2}+
      h_{m}^{1}(-1-m+\alpha )\right) }{2\, (m!)^{3}\,
    \Gamma \left( -1-m+\alpha \right) ^{3}}\, \xi ^{3}\nonumber \\
  &  & +\, \frac{\left( -1\right) ^{m}\, (1+m-\alpha )}
  {m!\, \Gamma (-1-m+\alpha )}\, \xi \, \zeta ^{2}\nonumber \\
  &  & +\, \frac{1}{6\, (m!)^{4}\, \Gamma (-1-m+\alpha )^{4}}\,
  \bigl (4\, h_{m}^{0}(-1-m+\alpha )\label{TRUN1} \\
  &  & \quad \times \left( 4\, h_{m}^{0}(-1-m+\alpha )^{2}+3\,
    h_{m}^{2}(-1-m+\alpha )\right) \nonumber \\
  &  & \quad +\, h_{m}^{2}(-1-m+\alpha )\bigr )\, \xi ^{4}\nonumber \\
  &  & +\, \frac{3-2\, (1+m-\alpha )\,
    h_{m}^{0}(-m+\alpha )}{(m!)^{2}\,
    \Gamma (-1-m+\alpha )^{2}}\, \xi ^{2}\zeta ^{2}+\cdots ,\nonumber
\end{eqnarray}
\begin{eqnarray}
  z_{2,m}(\xi ,\eta ,\zeta ) & = &
  -\alpha -m+\frac{\left( -1\right) ^{m+1}}{m!\,
    \Gamma (-m-\alpha )}\, \eta +\frac{h_{m}^{0}(-m-\alpha )}
  {(m!)^{2}\, \Gamma (-m-\alpha )^{2}}\, \eta ^{2}\nonumber \\
  &  & +\, \frac{\left( -1\right) ^{m+1}
    \left( 3\, h_{m}^{0}(-m-\alpha )^{2}+
      h_{m}^{1}(-m-\alpha )\right) }{2\, (m!)^{3}\,
    \Gamma \left( -m-\alpha \right) ^{3}}\, \eta ^{3}\nonumber \\
  &  & +\, \frac{\left( -1\right) ^{m}\, (m+1)}{m!\,
    \Gamma (-m-\alpha )}\, \eta \, \zeta ^{2}\nonumber \\
  &  & +\, \frac{1}{6\, (m!)^{4}\, \Gamma (-m-\alpha )^{4}}\,
  \bigl (4\, h_{m}^{0}(-m-\alpha )\label{TRUN2} \\
  &  & \quad \times \left( 4\, h_{m}^{0}(-m-\alpha )^{2}+3\,
    h_{m}^{2}(-m-\alpha )\right) \nonumber \\
  &  & \quad +\, h_{m}^{2}(-m-\alpha )\bigr )\, \eta ^{4}\nonumber \\
  &  & +\, \frac{1-2\, (m+1)\, h_{m}^{0}(-m-\alpha )}{(m!)^{2}\,
    \Gamma (-m-\alpha )^{2}}\, \eta ^{2}\zeta ^{2}+\cdots .\nonumber
\end{eqnarray}

A similar analysis can be carried out to get the asymptotic behaviour
of roots for \( \xi ,\eta ,\zeta  \) large. To this end assume that
\( \xi \eta -\zeta ^{2}\neq 0 \) and set
\[
\xi '=\frac{\xi }{\xi \eta -\zeta ^{2}}\, ,
\textrm{ }\eta '=\frac{\eta }{\xi \eta -\zeta ^{2}}\, ,
\textrm{ }\zeta '=\frac{\zeta }{\xi \eta -\zeta ^{2}}\, .
\]
Notice that \( \xi '\eta '-{\zeta '}^{2}=(\xi \eta -\zeta ^{2})^{-1} \).
Equation (\ref{EV}) becomes
\begin{equation}
  \label{EVINF}
  \frac{\xi '\eta '-{\zeta '}^{2}}{\Gamma (z)\,
    \Gamma (z+\alpha )}+\frac{\xi '}{\Gamma (z+\alpha -1)\,
    \Gamma (z+\alpha )}+\frac{\eta '}{\Gamma (z)^{2}}+
  \frac{1}{\Gamma (z)\, \Gamma (z+\alpha -1)}=0.
\end{equation}
Roots of (\ref{EVINF}) are analytic functions in
\( \xi ',\eta ',\zeta ' \)
at least in some neighbourhood of the origin. Again, it would be possible
to compute the beginning of the corresponding power series and to
derive formulae similar to those of (\ref{TRUN1}), (\ref{TRUN2})
but we avoid doing it here explicitly.

Instead we prefer to plot two graphs in order to give a reader some
impression about how the eigenvalues may depend on the parameters,
i.e. on the boundary conditions. In each graph we choose a line in
the parameter space,
\( \{(\xi t ,\eta t,\zeta t)\in \R ^{3};\  t\in \R\}\),
and we depict the dependence on $t$ of several first eigenvalues
for the corresponding self-adjoint extension restricted
to \( \HH _{c} \) (see (\ref{SUBS}) for the substitution).
In the both graphs we have set $\alpha=0.3$ and $B=1$.

Probably the most complete general information which is available
about solutions of equation (\ref{EV}) might be a localisation of
roots of this equation with respect to a suitable splitting of the
real line into intervals. Let us choose the splitting into intervals
with boundary points coinciding with the roots of equation (\ref{RINF}).
To get the localisation let us rewrite equation (\ref{EV}), equivalently
provided \( z\neq -\Z _{+}\cup (1-\alpha -\Z _{+}) \), as follows

\begin{equation}
  \label{EVLOC}
  \left( \frac{\Gamma (z-1+\alpha )}{\Gamma (z)}+
    \xi \right) \left( \frac{\Gamma (z)}{\Gamma (z+\alpha )}+
    \eta \right) =\zeta ^{2}.
\end{equation}
Put
\[
F_{\alpha }(z)=\frac{\Gamma (z-1+\alpha )}{\Gamma (z)}
\]
so that equation (\ref{EVLOC}) can be rewritten as
\begin{equation}
  \label{EVF}
  \left( F_{\alpha }(z)+\xi \right)
  \left( F_{1-\alpha }(z+\alpha )+\eta \right) =\zeta ^{2}.
\end{equation}

It is easy to carry out some basic analysis of the function
\( F_{\alpha }(z) \).
We have
\( {F_{\alpha }}'(z)=F_{\alpha }(z)\, (\psi (z-1+\alpha )-\psi (z)) \).
One observes that
\( F_{\alpha }(z)>0 \) for \( z\in {\, ]1-\alpha ,+
\infty [\, }\, \cup \left( \bigcup _{m\in \Z _{+}}\, {\, ]-
  \alpha -m,-m[\, }\right)  \),
and \( F_{\alpha }(z)<0 \) for
\( z\in \bigcup _{m\in \Z _{+}}\, {\, ]-m,1-\alpha -m[\, } \),
and in any case \( F_{\alpha }{}'(z)<0 \). In the former case this
follows from the fact that \( \psi (z) \) is strictly increasing
on each of the intervals
\( {\, ]0,+\infty [\, } \) and \( {\, ]-m-1,-m[\, } \),
with \( m\in \Z _{+} \). In the latter case this is a consequence
of the identity
\[
\psi (z-1+\alpha )-\psi (z)=
\frac{\pi \, \sin (\pi \alpha )}{\sin (\pi z)\,
  \sin (\pi (z+\alpha))}+
\int _{0}^{\infty }\frac{e^{-\left( 1-z\right) \, t}
  \left( 1-e^{-\left( 1-\alpha \right) \, t}\right) }{1-e^{-t}}\, \dd
t.
\]
Moreover,
\[
\lim _{z\rightarrow +\infty }F_{\alpha }(z)=0,
\textrm{ }\lim _{z\rightarrow (1-\alpha -m)\pm }
F_{\alpha }(z)=\pm \infty \textrm{ and }F_{\alpha }(-m)=0
\textrm{ for }m\in \Z _{+}.
\]
This also implies that \( F_{1-\alpha }(z+\alpha )>0 \) for
\( z\in {\, ]0,+\infty [\, }\, \cup \,
\left( \bigcup _{m\in \Z _{+}}\, {\, ]-1-m,-\alpha -m[\, }\right)  \)
and \( F_{1-\alpha }(z)<0 \) for
\( z\in \bigcup _{m\in \Z _{+}}\, {\, ]-\alpha -m,-m[\, } \),
in any case \( F_{1-\alpha }{}'(z+\alpha )<0 \), and
\begin{eqnarray*}
  &  & \lim _{z\rightarrow +\infty }F_{1-\alpha }(z+\alpha )=0,
  \textrm{ }\lim _{z\rightarrow -m\pm }F_{1-\alpha }(z+\alpha )=
  \pm \infty ,\\
  &  & \textrm{and }F_{1-\alpha }(-\alpha -m)=0
  \textrm{ for }m\in \Z _{+}.
\end{eqnarray*}

With the knowledge of these basic properties of the function
\( F_{\alpha }(z) \)
it is a matter of an elementary analysis to determine the number of
roots of equation (\ref{EVF}) in each of the intervals
\( {\, ]1-\alpha ,+\infty [\, } \),
\( {\, ]-m,1-\alpha -m[\, } \) and \( {\, ]-\alpha -m,-m[\, } \),
with \( m\in \Z _{+} \). The result is summarised in the following
tables.\vskip 6pt

{\centering
  \begin{tabular}{|c|c|c||c|}
    \hline
    \multicolumn{4}{|c|}{interval \( \, ]1-\alpha ,+\infty [\,  \)}\\
    \hline
    \multicolumn{3}{|c||}{conditions}&
    number of roots\\
    \hline
    \hline
    \( \xi \geq 0 \)&
    \( \eta \geq 0 \)&
    \( \zeta ^{2}>\xi \eta  \)&
    1\\
    \hline
    \( \xi \geq 0 \)&
    \( \eta \geq 0 \)&
    \( \zeta ^{2}\leq \xi \eta  \)&
    0\\
    \hline
    \( \xi \geq 0 \)&
    \( -\Gamma (1-\alpha )<\eta <0 \)&
    no condition&
    1\\
    \hline
    \( \xi \geq 0 \)&
    \( \eta \leq -\Gamma (1-\alpha ) \)&
    no condition&
    0\\
    \hline
    \( \xi <0 \)&
    \( \eta \geq 0 \)&
    no condition&
    1\\
    \hline
    \( \xi <0 \)&
    \( -\Gamma (1-\alpha )<\eta <0 \)&
    \( \zeta ^{2}\geq \xi \eta  \)&
    1\\
    \hline
    \( \xi <0 \)&
    \( -\Gamma (1-\alpha )<\eta <0 \)&
    \( \zeta ^{2}<\xi \eta  \)&
    2\\
    \hline
    \( \xi <0 \)&
    \( \eta \leq -\Gamma (1-\alpha ) \)&
    \( \zeta ^{2}\geq \xi \eta  \)&
    0\\
    \hline
    \( \xi <0 \)&
    \( \eta \leq -\Gamma (1-\alpha ) \)&
    \( \zeta ^{2}<\xi \eta  \)&
    1\\
    \hline
  \end{tabular}\par}

\vskip 6pt

{\centering
  \begin{tabular}{|c|c||c|}
    \hline
    \multicolumn{3}{|p{18em}|}
    {\centering interval \( \, ]0,1-\alpha [\,  \)}\\
    \hline
    \multicolumn{2}{|p{10em}||}{\centering conditions}&
    number of roots\\
    \hline
    \hline
    \( \xi \leq 0 \)&
    \( \eta \geq -\Gamma (1-\alpha ) \)&
    0\\
    \hline
    \( \xi \leq 0 \)&
    \( \eta <-\Gamma (1-\alpha ) \)&
    1\\
    \hline
    \( \xi >0 \)&
    \( \eta \geq -\Gamma (1-\alpha ) \)&
    1\\
    \hline
    \( \xi >0 \)&
    \( \eta <-\Gamma (1-\alpha ) \)&
    2\\
    \hline
  \end{tabular}\par}

\vskip 6pt

{\centering
  \begin{tabular}{|c|c||c|}
    \hline
    \multicolumn{3}{|p{18em}|}
    {\centering intervals \( \, ]-\alpha -m,-m[\,  \),
      \( m\in \Z _{+} \)}\\
    \hline
    \multicolumn{2}{|p{10em}||}{\centering conditions}&
    number of roots\\
    \hline
    \hline
    \multicolumn{1}{|p{5em}|}{\centering \( \xi \geq 0 \)}&
    \( \eta \leq 0 \)&
    0\\
    \hline
    \( \xi \geq 0 \)&
    \( \eta >0 \)&
    1\\
    \hline
    \( \xi <0 \)&
    \( \eta \leq 0 \)&
    1\\
    \hline
    \( \xi <0 \)&
    \( \eta >0 \)&
    2\\
    \hline
  \end{tabular}\par}

\vskip 6pt

{\centering
  \begin{tabular}{|c|c||c|}
    \hline
    \multicolumn{3}{|p{18em}|}
    {\centering intervals \( \, ]-1-m,-\alpha -m[\,  \),
      \( m\in \Z _{+} \)}\\
    \hline
    \multicolumn{2}{|p{10em}||}{\centering conditions}&
    number of roots\\
    \hline
    \hline
    \multicolumn{1}{|p{5em}|}{\centering \( \xi \leq 0 \)}&
    \( \eta \geq 0 \)&
    0\\
    \hline
    \( \xi \leq 0 \)&
    \( \eta <0 \)&
    1\\
    \hline
    \( \xi >0 \)&
    \( \eta \geq 0 \)&
    1\\
    \hline
    \( \xi >0 \)&
    \( \eta <0 \)&
    2\\
    \hline
  \end{tabular}\vskip 6pt\par}

\noindent This is to be completed with the simple observation that
\( 1-\alpha  \) is a root of (\ref{EV}) if and only if
\( \eta =-\Gamma (1-\alpha ) \),
and \( -m \), with \( m\in \Z _{+} \), is a root if and only if
\( \xi =0 \), and finally \( -\alpha -m \), with \( m\in \Z _{+} \),
is a root if and only if \( \eta =0 \).

Let us note that this localisation is in agreement with a general
result according to which if \( A \) and \( B \) are two self-adjoint
extensions of the same symmetric operator with finite deficiency indices
\( (d,d) \) then any interval \( J\subset \R  \) not intersecting
the spectrum of \( A \) contains at most \( d \) eigenvalues of
the operator \( B \) (including multiplicities) and no other part
of the spectrum of \( B \) \cite[\S 8.3]{Weidmann}. Thus in our
example if \( J \) is an open interval whose boundary points are
either two subsequent eigenvalues of \( H^{\infty } \) or the lowest
eigenvalue of \( H^{\infty } \) and \( -\infty  \) then any self-adjoint
extension \( H^{\Lambda } \) has at most two eigenvalues in \( J \).

\section{Concluding remarks}

The above discussion does not exhaust all questions related to the
system under consideration. One may ask, for instance, how the
state of such a particle evolves under an adiabatic change of
parameters. In particular, since the model exhibits eigenvalue
crossings, one may expect that there are parameter loops
exhibiting a nontrivial Berry phase. Another question concerns the
physical meaning of our idealized model. More specifically, one is
interested in which sense the model Hamiltonian can be
approximated by those with smeared flux and a regular interaction.
We leave these problems to a future publication.

\newpage

\begin{figure}[h]
{\centering \hspace*{-35mm}\includegraphics{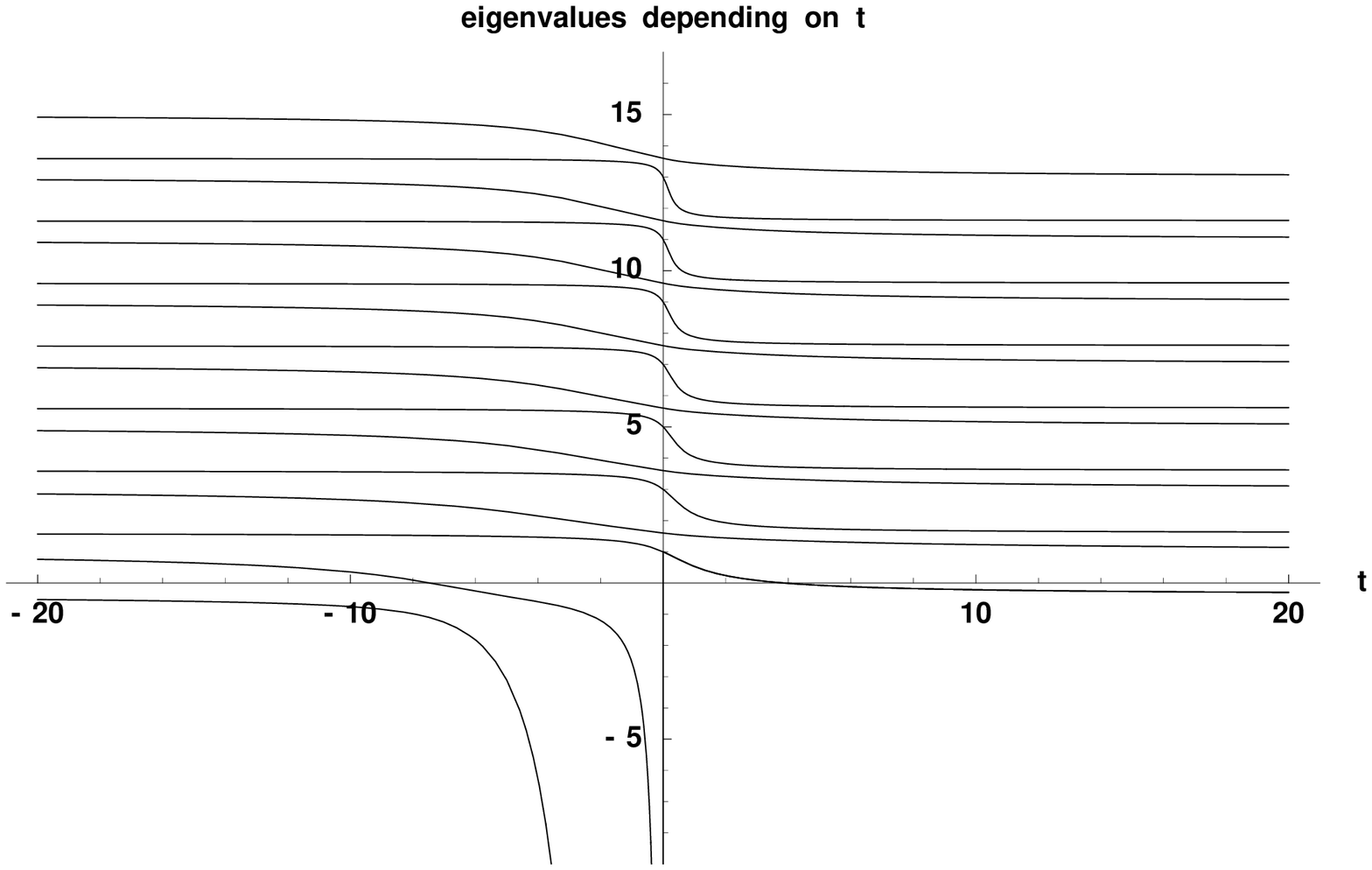} \par}
\vskip 116pt
\caption{The Hamiltonian is determined by the boundary conditions
  corresponding to the parameters
  $(\xi,\eta,\zeta)=(0.95\,t,0.25\,t,0.25\,t)$, $\alpha=0.3$, $B=1$.}
\end{figure}
\newpage

\begin{figure}[h]
{\centering \hspace*{-35mm}\includegraphics{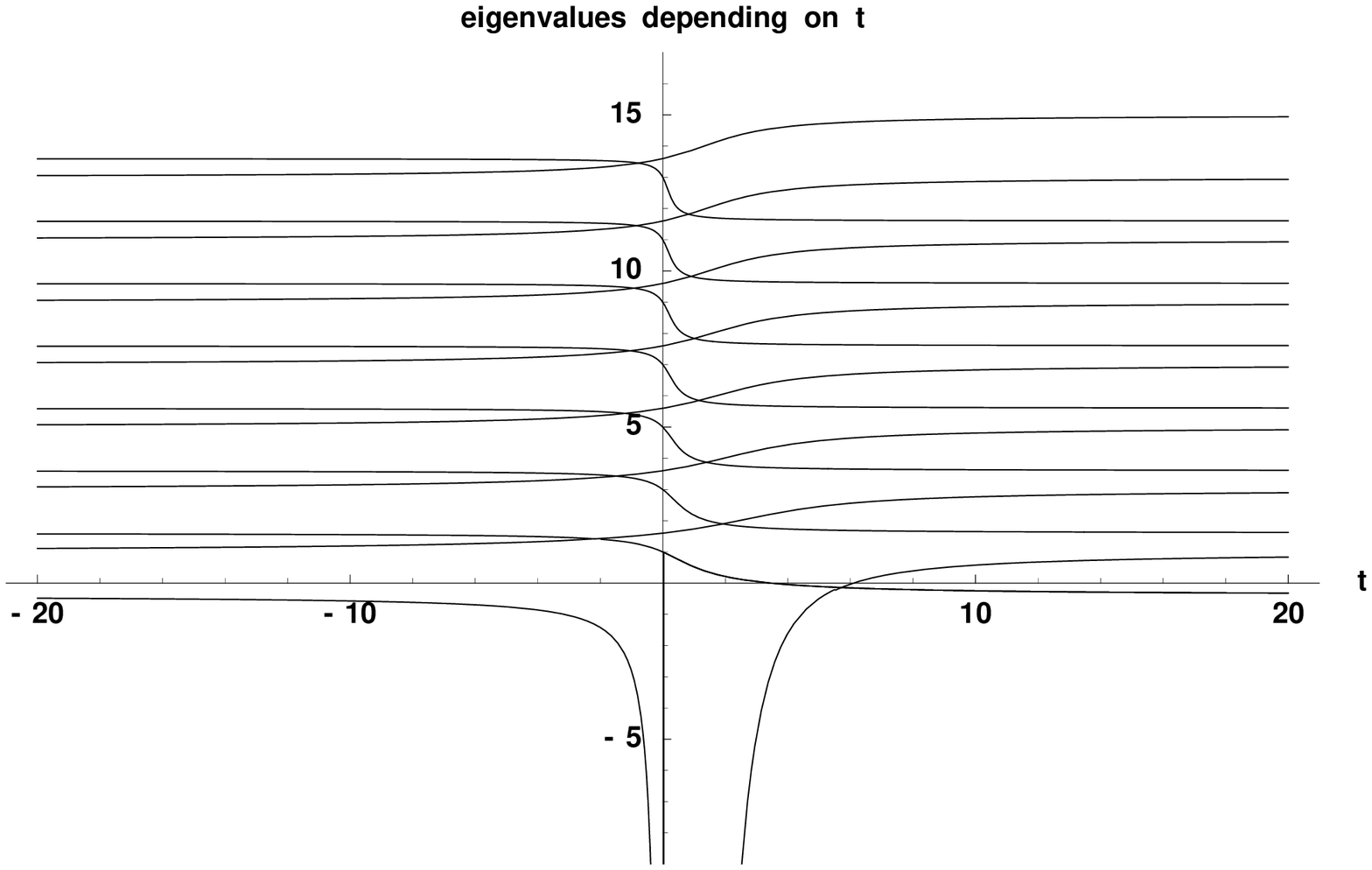} \par}
\vskip 116pt
\caption{The Hamiltonian is determined by the boundary conditions
  corresponding to the parameters
  $(\xi,\eta,\zeta)=(0.95\,t,-0.25\,t,0)$, $\alpha=0.3$, $B=1$.}
\end{figure}
\newpage

\vskip 24pt \noindent{\bf Acknowledgements.} The research was
partially supported by Grants GACR 201/01/01308 and GAAS 1018101.

\end{document}